# A nonextensive approach to Bose-Einstein condensation of trapped interacting boson gas

A. Lawani[a], J. Le Meur[a], Dmitrii Tayurskii[b], A. El Kaabouchi[a], L. Nivanen[a], B. Minisini[a], F. Tsobnang[a], M. Pezeril[c], A. Le Méhauté[a], Q. A. Wang[a]

[a]Institut Supérieur des Matériaux et Mécaniques Avancés du Mans, 44, Avenue F.A. Bartholdi, 72000 Le Mans, France

[b]Departement of physics, Kazan State University, Kazan 420008, Russia

[c]Faculté de Sciences et Techniques, Université du Maine, 72000 Le Mans, France

**Abstract**

In the Bose-Einstein condensation of interacting atoms or molecules such as $^{87}Rb$, $^{23}Na$ and $^{7}Li$, the theoretical understanding of the transition temperature is not always obvious due to the interactions or zero point energy which cannot be exactly taken into account. The S-wave collision model fails sometimes to account for the condensation temperatures. In this work, we look at the problem within the nonextensive statistics which is considered as a possible theory describing interacting systems. The generalized energy $U_q$ and the particle number $N_q$ of boson gas are given in terms of the nonextensive parameter $q$. $q>1$ ($q<1$) implies repulsive (attractive) interaction with respect to the perfect gas. The generalized condensation temperature $T_{cq}$ is derived versus $T_c$ given by the perfect gas theory. Thanks to the observed condensation temperatures, we find $q \approx 0.1$ for $^{87}Rb$ atomic gas, $q \approx 0.95$ for $^{7}Li$ and $q \approx 0.62$ for $^{23}Na$. It is concluded that the effective interactions are essentially attractive for the three considered atoms, which is consistent with the observed temperatures higher than those predicted by the conventional theory.

Since 1995, the creation of Bose-Einstein condensation (BEC) in dilute atomic gases of $^{87}Rb$ [1], $^{7}Li$ [2] and $^{23}Na$ [3] and others has stimulated a great deal of interest in the statistical investigation of interacting (imperfect) quantum particles systems. It is realized that the conventional Bose-Einstein statistics (BES) sometimes fails to yield the observed value of the transition temperature $T_c$. A good example is $^{4}He$ for which the observed transition temperature is $T_c = 2.17K$ and the theoretical one is $T_c = 3.10K$. For the dilute atomic gases trapped in harmonic potential mentioned above, there are also significant differences between the observed and theoretical $T_c$ [1][2][3]. In addition, the non interacting gas picture gives an atom velocity distribution of the condensate which is not consistent with that observed in the atomic vapour [1][3].

These failures are obviously due to the interaction between the particles neglected in BES. One of the treatments taking into account this interaction is proposed by Huang [4] with following approximations at very low temperature: 1) the particle interaction takes place through binary collision (short term interaction); 2) the effective interaction is weak; 3) a particle sees only average effect of the interaction (mean field); 4) only first order perturbation is considered. On this basis, Huang has given the interaction energy [4]

$$\delta U \approx \frac{a 4\pi\hbar^2}{m}$$ where $a$ is called *scattering length of S-waves* of the atomic collision and $m$ is the particle mass. In general, $a$ is positive for predominantly repulsive interaction and negative for predominantly attractive one. For $^{87}Rb$ atomic vapour at low temperature, $a = 200a_0$ (repulsive) [1], for $^{23}Na$ gas, $a = (92 \pm 25)a_0$ (repulsive) [3] and for $^{7}Li$ gas, $a = (-27\pm0.8)a_0$ (attractive) [2] ($a_0$ is the Bohr radius). This theory has problem with $^{7}Li$ gas because $a < 0$ leads to imaginary physical quantities of the gas [2][3] making the BEC of $^{7}Li$ dilute gas impossible.

In this work, we try to study the interacting quantum gas within nonextensive statistical mechanics (NSM) [5][6][7] which is considered as a possible theory for interacting system. Within NSM, a boson distribution is given by[8]

$$\langle n_q \rangle = \frac{1}{[1+(q-1)\frac{\varepsilon - \mu}{kT}]^{\frac{1}{q-1}} - 1} \tag{1}$$

where $n_q$ is the average occupation number at a state with energy $\varepsilon$ and chemical potential $\mu$ and the parameter $q$ is a positive real number. NSM recovers BES when $q = 1$. In the $q \neq 1$ case, the internal energy of the system varies as $q$ changes. According to previous

results[9][10], the interactions are repulsive (or attractive) for $q > 1$ (or $q < 1$). In what follows, this formalism will be applied to the dilute gases of $^{87}Rb$, $^{7}Li$ and $^{23}Na$ atoms trapped in harmonic potentials reported in [1][2][3].

For imperfect boson gas trapped in a three-dimensional harmonic potential

$$V(X, Y, Z) = \frac{1}{2}K(X^2 + Y^2 + Z^2) = \frac{1}{2}KR^2, \quad (2)$$

the total number of particles of mass $m$ is given by

$$N = \frac{2(4\pi)^2}{h^3}(\frac{m}{K})^{\frac{3}{2}}(kT)^3 \int_0^{x_q} \int_0^{y_q} \frac{x^{1/2}dx y^{1/2}dy}{[1+(q-1)(x+y-\upsilon)]^{\frac{1}{q-1}}-1} \quad (3)$$

where $x = \frac{p^2}{2mkT}$, $y = \frac{KR^2}{2kT}$ and $\upsilon = \frac{\mu}{kT}$. For $q < 1$, $x_q + y_q = \frac{q}{1-q} + \upsilon$ due to Tsallis cut-off condition [5]. For $q > 1$, $x_q = \infty$ and $y_q$ depends on the living-space of the particles. If we let the quantum mechanical living space tend to infinity, then $y_q \to \infty$ [11]. However, for large $x$ and $y$, $q$ must be limited in order that the integral (3) converges. For this purpose, we write the integrand as $\frac{x^{1/2}y^{1/2}}{(x+y)^{\frac{1}{q-1}}} = \frac{x^{1/2}}{(x+y)^{\frac{1}{2(q-1)}}} \frac{y^{1/2}}{(x+y)^{\frac{1}{2(q-1)}}} \leq \frac{x^{1/2}}{x^{\frac{1}{2(q-1)}}} \frac{y^{1/2}}{y^{\frac{1}{2(q-1)}}}$ for large $x$ and $y$. This leads to $q < \frac{4}{3}$ for the convergence of the integral (note that $q < \frac{5}{3}$ for free particle model [9][10]). Eq.(3) can be written as follows:

$$N = N_0 + QI_qN(0) \quad (4)$$

where $N$ is the total trapped particle number and

$$N_0 = \frac{1}{[1+(q-1)(\frac{3}{2}\hbar\omega - \upsilon)]^{\frac{1}{q-1}}-1} \quad (5)$$

is the occupation number of the ground state with $\varepsilon = \frac{3}{2}\hbar\omega$, $Q = (\frac{kT}{\hbar})^3(\frac{m}{K})^{\frac{3}{2}} = (\frac{kT}{\hbar\omega})^3$ and

$$I_qN(\upsilon) = \frac{4}{\pi}\int_0^{x_q}\int_0^{y_q}\frac{x^{1/2}dx y^{1/2}dy}{[1+(q-1)(x+y-\upsilon)]^{\frac{1}{q-1}}-1} \quad (6)$$

The critical temperature $T_{cq}$ of Bose-Einstein condensation is defined by

$$I_qN(0) = \frac{N}{Q} \quad (7)$$

So that

$$kT_{cq} = \hbar\omega(\frac{N}{I_q N(0)})^{1/3}. \qquad (8)$$

where ω is the average trapping frequency, $\hbar$ the Planck constant and $k$ the Boltzmann constant. The variation of the integral $I_q N(0)$ with increasing $q$ is given in Figure 1.

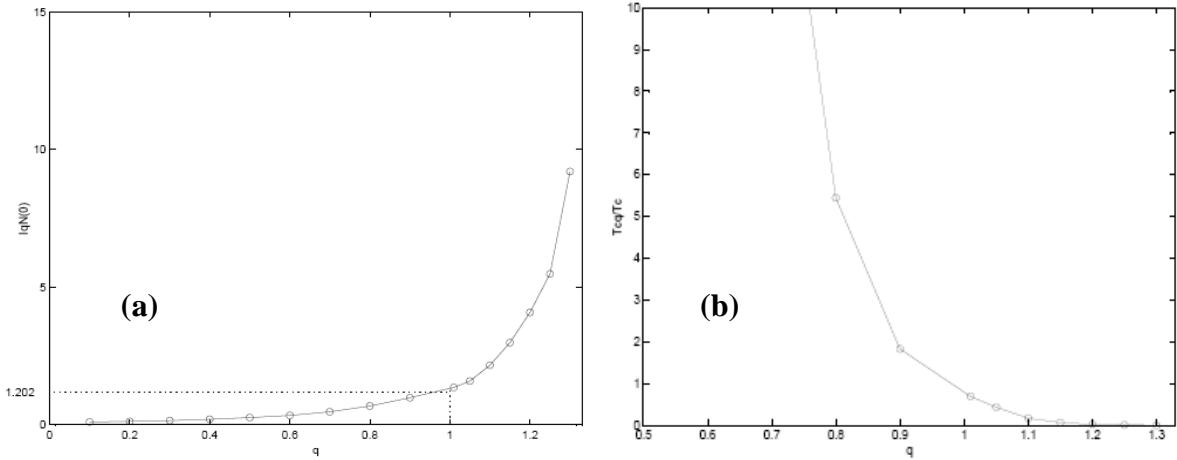

**Figure 1**, (a) q variation of integral $I_q N(0) = \frac{4}{\pi}\int_0^{x_q}\int_0^{y_q} \frac{x^{1/2}dxy^{1/2}dy}{[1+(q-1)(x+y)]^{\frac{1}{q-1}}-1}$.
Note that for q=1, $I_1 N(0) = 1.202$ and $I_q N(0)$ diverges for q=1.333. **(b)** q variation of the ratio $T_{cq}/T_c = [1.202/I_q N(0)]^3$. The ratio is zero for the limit q=1.333.

If $T < T_{cq}$ (μ = 0), we can write:

$$\frac{N_0}{N} = 1 - \frac{I_q N(0)}{\alpha} \qquad (9)$$

where α is given by:

$$\alpha = \frac{N}{Q} = N\left(\frac{\hbar\omega}{kT}\right)^3 \qquad (10)$$

It is straightforward to write

$$\frac{N_0}{N} = 1 - \left(\frac{T}{T_{cq}}\right)^3 \qquad (11)$$

which gives the percentage of degenerated (condensed) particles when $T < T_{cq}$. In the case of $q = 1$, Eq.(7) gives the conventional result of BEC:

$$I_1(0) = N(\frac{\hbar\omega}{kT_c})^3 = 1.202 \quad or \quad T_c = \frac{\hbar\omega}{k}\left(\frac{N}{1.202}\right)^{1/3} \tag{12}$$

From Eqs.(7) and (12), we can straightforwardly find the relation between the generalized critical temperature $T_{cq}$ and the conventional one $T_c$:

$$T_{cq} = T_c\left[\frac{1.202}{I_q(0)}\right]^{1/3} \tag{13}$$

From the $q$-dependence of $Iq(0)$ shown in Figure 1a, we can calculate the variation of $T_{cq}$ or $T_{cq}/Tc$ with increasing $q$. This $T_{cq}$ behaviour is plotted in Figure 1b. Note that $T_{cq} = 0$ for $q = \frac{4}{3} = 1.333$. Note that Eq.(11) can be written as

$$\frac{N_0}{N} = 1 - \frac{IqN(0)}{1.202}\left(\frac{T}{T_c}\right)^3 \tag{14}$$

Which is plotted in figure 2.

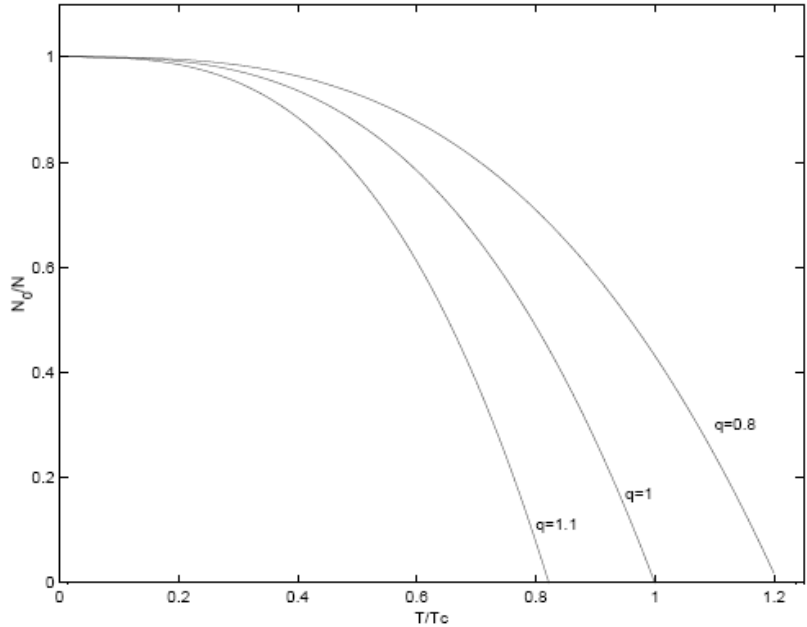

**Figure 2**, The variation of the ratio $\frac{N_0}{N} = 1 - \frac{IqN(0)}{1.202}\left(\frac{T}{T_c}\right)^3$ with increasing $T/T_c$ for 3 different values of $q$.

Now let us consider the internal energy of the systems

$$U_q = \sum_i \varepsilon_i \langle n_q \rangle \quad (15)$$

$$= \frac{1}{h^3} \int_0^{P_q} \int_0^{R_q} \frac{(\frac{p^2}{2m}+\frac{KR^2}{2})4\pi p^2 dp 4\pi R^2 dR}{[1+(q-1)(\frac{p^2}{2m}+\frac{KR^2}{2}-\upsilon)]^{\frac{q}{q-1}}-1}$$

With the same variable changes as in Eq. (**3**), Eq. (15) can be recast as

$$U_q = (\frac{kT}{\hbar\omega})^3 (kT) \frac{4}{\pi} \int_0^{x_q} \int_0^{y_q} \frac{(x+y)x^{1/2}dxy^{1/2}dy}{[1+(q-1)(x+y-\upsilon)]^{\frac{q}{q-1}}-1} \quad (16)$$

$$= QI_q U(\upsilon)kT$$

where

$$I_q U(\nu) = \frac{4}{\pi} \int_0^{x_q} \int_0^{y_q} \frac{(x+y)x^{1/2}dxy^{1/2}dy}{[1+(q-1)(x+y-\nu)]^{\frac{q}{q-1}}-1} \quad (17)$$

From Eq. (7), Eq.(16) can be written as follows:

$$\frac{U_q}{NkT} = \frac{I_q U(\upsilon)}{I_q N(\upsilon)} \quad (18)$$

With the same method as for Eq.(3), we can see that $U_q$ diverges for $q \geq 6/5$, i.e. the system is no longer stable due to the strong repulsive interaction represented by $q$ larger than unity.

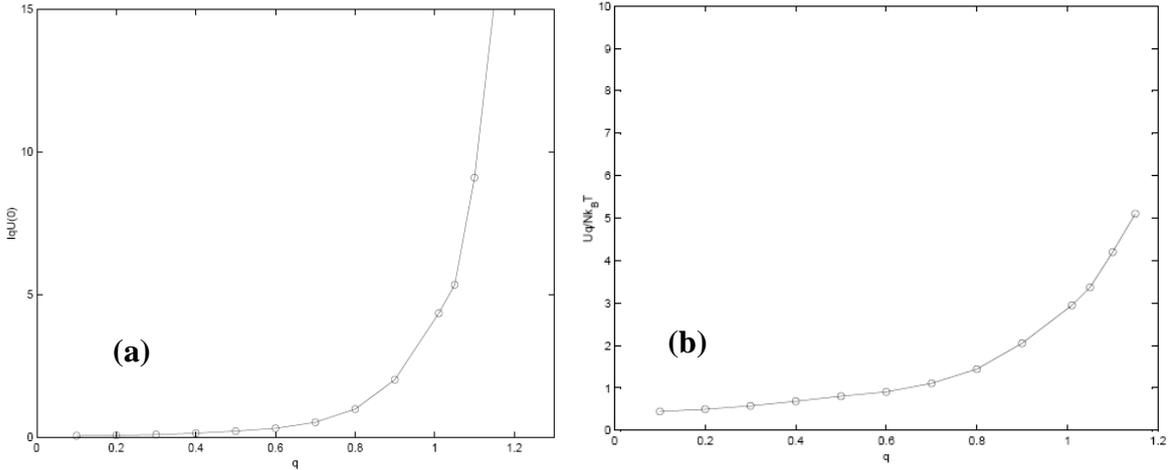

**Figure 3**, (**a**) q variation of integral $I_q U(0) = \frac{4}{\pi} \int_0^{x_q} \int_0^{y_q} \frac{(x+y)x^{1/2}dxy^{1/2}dy}{[1+(q-1)(x+y)]^{1/(q-1)}-1}$. Note that $I_q U(0)$ diverges for $q=6/5$. (**b**) q variation of the condensate energy $U_q/NkT = IqU(0)/IqN(0)$.

Numerical evaluation shows that $I_q U(0)$ for $T < T_{cq}$ is a monotonic increasing function of $q$ up to $q = 1.2$ at which it diverges (see figure 3a). When $q = 1$, $I_1 U(0) = 3.25$ and $U_1 = 3.25 Q k T = 2.7 N k T$ (conventional value). The numerical result of Eq.(18) is plotted in Figure 2b for $\upsilon = 0$. It is worth noticing in figure (3b) that $U_{q \leq 1} \leq 2.7 N k T \leq U_{q \geq 1}$. This means that, with respect to the conventional condensate ($U_1 = 2.7 N k T$), the neglected interaction is repulsive for $q > 1$ and attractive for $q < 1$. This point of view will help us to understand the differences between the observed transition temperatures and theoretical ones given by the conventional BES. In what follows, the $q$-variation of $T_{cq}$ can be used to find the experimental value of the BEC temperature.

For $^{87}Rb$ gas, the theoretical $T_c = 74 nK$ (calculated from Eq.(12) with $N_c = 2 \times 10^4$ observed at the transition). Let $T_{cq} = 170 nK$, the observed condensation temperature[1], from Eq.(13), we obtain $I_q N(0) = 1.202/2.297^3 = 0.0991$ corresponding to $q \approx 0.1$. This means that the interaction represented by $q$ is attractive making it possible to observe the condensation at a higher temperature than the theoretical prediction without interaction.

For $^{23}Na$ gas, $T_c = 1350 nK$ (calculated with observed $Nc = 15 \times 10^6$ and $\omega = 2\pi \times 123$ Hz) and $T_{cq} = 2000 nK$ (observed temperature[3]), which yields $I_q N(0) = 1.202/1.482^3 = 0.37$ corresponding to $q \approx 0.62$, and implies an attractive interaction as well.

For $^7Li$ gas, $T_c = 386 nK$ (calculated with observed $N_c = 2 \times 10^5$ and $\omega = 2\pi \times 146$ Hz) and $T_{cq} = 400 nK$ (observed temperature)[2] yield $I_q N(0) = 1.202/1.036^3 = 1.08$ and $q \approx 0.95$ implying a weak attractive interaction.

The conclusion of this work is:

1) The three observed transition temperatures are all higher than the theoretically predicted values. So it turns out that the empirical values of $q$ are all smaller than unity, which means there are attractive interatomic interactions in the three condensates. These attractions make the particles easier to be condensed at higher temperatures than those predicted by BES.

2) Above interpretation seems natural and coherent, but it is in conflict with Huang's theory for $^{87}Rb$ and $^{23}Na$ because these gases have positive scattering length $a$ corresponding to repulsive interaction with $\delta U > 0$. In our opinion, repulsive interaction would make the particles more difficult to be condensed hence need a lower transition temperature than predicted. This is not the observed cases. However, considering the experimental uncertainty and the crude approximation in the theory of Huang, we perhaps have to wait for future experimental results on other properties of these condensates which may shed light on this doubt, i.e., attractive or repulsive interaction in the condensates which have higher $T_c$ than that predicted by BES.

3) This work makes it possible to give a clear physical content to the parameter *q* of NSM which can be used to represent neglected interactions in the treatment of perfect gas like model of NSM. However, in view of the uncertainty of the experimental measurements, the q values found in this work are also subject to uncertainty. Further study with more precise experimental results would be useful to confirm the present work.